\documentstyle[pre,aps,epsf]{revtex}

\newcommand{\bq}{\begin{equation}}
\newcommand{\ee}{\end{equation}}
\newcommand{\fr}[2]{\frac{#1}{#2}}

\begin{document}
\draft


\title{Summing graphs for random band matrices}

\author{P.G.Silvestrov}

\address{Budker Institute of Nuclear Physics, 630090 Novosibirsk, Russia}

\address{The Niels Bohr Institute,
Blegdamsvej 17, DK-2100 Copenhagen {\O}, Denmark}

\maketitle

\begin{abstract}
A method of resummation of infinite series of perturbation theory
diagrams is applied for studying the properties of random band
matrices. The topological classification of Feynman diagrams, which
was actively used in last years for matrix model regularization of
$2d$-gravity, turns out to be very useful for band matrices.  The
critical behavior at the edge of spectrum and the asymptotics of
energy level correlation function are considered. This correlation
function together with the hypothesis about universality of spectral
correlations allows to estimate easily the localization length for
eigen-vectors. A smoothed two-point correlation function of local
density of states $\overline{ \rho(E_1,i) \rho(E_2,j)_c}$ as well as
the energy level correlation for finite size band matrices are also
found. As $d$-dimensional generalization of band matrices lattice
Hamiltonians with long-range random hopping are considered as well.

\end{abstract}


\section{Introduction}\label{sec:1}

Random band matrices were introduced many years ago by E.Wigner
\cite{Wigner} as a model Hamiltonian for complicated quantum
systems. In the last few years statistical properties of random band
matrices have become again the subject of intensive analytical and
numerical investigation \cite{Mirlin,Pichard,Izrailev} due to their
application to condensed matter physics and statistics of spectrum of
chaotic systems.

Up to now all the analytical results for this quasi-$1d$ systems (for
review see \cite{Mirlin}) were obtained by mapping them onto a
super-symmetric $\sigma$-model \cite{efetov}. However, in this paper
we would like to develop another method for calculations with random
banded matrix ensembles. Roughly speaking our method consists of
summation of infinite series of perturbation theory
diagrams. Diagrammatic methods were used many years ago
\cite{Verb1,Pastur} for investigation of Gaussian ensembles of
$N\times N$ matrices but later this approach was almost forgotten for
years. Our aim in this paper will be to show how this ``ancient''
method may lead rather easily to new results for band matrices.

Let us consider a Gaussian ensemble of random band matrices.  Due to
the Wick theorem such ensembles may be completely defined by the
second moment:
\bq\label{band} 
\overline{H_{ij}H_{kn}} =
F(i-j)(\delta_{jk}\delta_{in} + 
\Theta \delta_{jn}\delta_{ik}) \,\, .
\ee 
The function $F(i-j)=F(|i-j|)$ vanishes very rapidly outside the band
(at $|i-j|>b\gg 1$).  Parameter $\Theta$ takes values $0$ or $1$. If
$\Theta=0$, one is dealing with Hermitian matrices of general form
(Gaussian unitary ensemble - GUE\footnote{Although, the notations GUE
and GOE for our ensembles are mainly the traditional since
unitary(orthogonal) invariance is broken explicitly for banded
matrices.}), while $\Theta=1$ corresponds to the real symmetric
matrices (Gaussian orthogonal ensemble - GOE). It is convenient to
define the width of the band $b$ and the typical strength of the
interaction $V$ through moments of the function $F$:
\begin{eqnarray}\label{defb}
&\,& b^2=\fr{F_2}{F_0} = \fr{\sum n^2 F(n)}{\sum F(n)} \\
&\,& V^2 = F_0 =\sum F(n)  \ \ \ . \nonumber
\end{eqnarray}
For practical computations we will sometimes use $F$ of the form
\begin{eqnarray}\label{F}
F(i-j)= \fr{V^2}{b\sqrt{2\pi}} \exp \left(
-\fr{(i-j)^2}{2b^2} \right) \,\, .
\end{eqnarray}
As will be shown below the results essentially do not depend on the
details of the shape of function $F(i-j)$.  We need only $F(i-j)$ to
be sufficiently smooth so that after averaging (\ref{band}) all
discrete sums may be replaced by integrals up to negligible
corrections $\sim \exp(-b)$. In doing so we still are able to consider
the corrections of any finite order in $1/b$.  Moreover, we argue
below that even smoothness of $F$ does not seem to be necessary for
most interesting applications (see (\ref{recur},\ref{initial}) and
discussion below, it will be shown also why (\ref{F}) is the most
natural choice of $F$).

By simple $d$-dimensional extension of band matrices one obtains
the Hamiltonian for particle hopping on a $d$-dimensional lattice
with random nonlocal interaction. This lattice model may be described
by the same formulas (\ref{band}-\ref{F}) with simple replacement of
all integer indices by the integer $d$-vectors and trivial
redefinition of~$F$
\bq\label{lattice}
i\rightarrow \vec{i} \,\,\, , \,\,\,
F(\vec{i} -\vec{j} ) = \fr{V^2}{b(2\pi)^{d/2}} \exp 
\left( -\fr{(\vec{i} -\vec{j} )^2}{2b^{2/d}} 
\right) \,\, .
\ee
Contrary to band matrices, properties of this model seem to be
completely unknown. The $b\gg 1$ in (\ref{lattice}) is effectively the
number of ``neighbors'', connected to each lattice site. The band
matrices (\ref{band}) may be associated now with the random
Hamiltonian for one dimensional lattice.  

The ``physical quantities'' which we would like to consider are
connected with the Green's function and the local density of states
\bq\label{Green}
G_{{i},{j}}(E) =
\left( \fr{1}{E-H}\right)_{{i},{j}} \,\,\, , \, \, \,
\rho(E,{i}) = \fr{1}{\pi} Im  
 G_{{i},{i}}(E-i0) \, \, \, .
\ee
There is no summation over $i$ in $G_{i,i}$ in the last formula. 
More specifically we would like to consider the averaged density
of states $\overline{\rho(E)}$ and the correlation of densities
for different but very close energies (and even for different
positions $i$ and $j$).

It is also constructive to compare our results
with those for the ensembles of usual $N\times N$ random
matrices, which are defined by the second moment
\bq\label{NN}
F(i-j) \equiv  \fr{V^2}{N} \, \, \, ; \, \, \,
\overline{H_{ij}H_{mn}} = \fr{V^2}{N} 
(\delta_{jm}\delta_{in} +
\Theta \delta_{jn}\delta_{im}) \,\, .
\ee
This Hamiltonian may be considered as the $d=0$ reduction of the
lattice model (\ref{lattice}) . Historically three main approaches
were applied for studying the statistics of the full $N\times N$
matrices.  The description of this approaches may be found e.g. in
papers \cite{Verb1,Verb2,Verb3}. The first one is summation of
infinite power series in $V/E$ \cite{Verb1}.  Another two methods are
the replica trick \cite{Verb2} and the super-symmetry method
\cite{Verb3}.

The success of both replicas and super-symmetry is essentially based
on the use of Hubbard--Stratonovich transformation. For $N\times N$
matrices this transformation reduces the problem to an almost trivial
calculation of a few dimensional integral. On the other hand for band
matrices even after Hubbard--Stratonovich one still is faced with a
$\sigma$-model on one-dimensional lattice. Thus it seems quite
probable that neither replicas, nor super-symmetry will lead to
considerable progress in the lattice model (\ref{lattice}).

The exact solution of $2d$ quantum gravity \cite{BKGMDS} stimulated
the explosion of interest in matrix models. In this application of
random matrices the discretized random surfaces appear as Feynman
graphs in the perturbative expansion of the matrix integral. However,
technically the famous double scaling solution of $2d$-gravity has
nothing to do with summation of graphs. For ensembles invariant under
orthogonal transformations it is useful to work with $N$ eigenvalues
instead of all $N^2$ matrix elements.  Unfortunately, for band
matrices, or lattice hopping Hamiltonian we could not find such a
simple solution, which is not based on the diagrammatic
expansion. Nevertheless for models (\ref{band},~\ref{lattice}) the
topological classification of diagrams, which arose in $2d$-gravity
(and originally in QCD \cite{tHooft}), simplifies drastically the
summation of the series.

This experience of dealing with the matrix integrals for $2d$-gravity
was later used for solving the problems typical for quantum chaos. In
\cite{AmbMak} a method of calculation of correlators of Green
functions for ensembles of large $N\times N$ matrices was
developed. An approach based on summation of perturbation theory
series for various ensembles of random matrices was also used in the
series of papers of Brezin and Zee (see e.g. \cite{BZ} and the later
papers of the same authors), though their diagrammatic technique
differs from used in the present paper.

The statistics of band matrices with parameters of the band slowly
varying along the diagonal was considered in \cite{Panleve}. The
behavior of the edge of spectrum for this extensions of the model
(\ref{band}) showed some surprising similarities with the edge
properties of matrix ensembles considered for the $2d$-gravity.  The
topological classification of the diagrams which we will explore below
was also briefly discussed in \cite{Panleve}.

The organization of this article is as follows. The general
description of the diagrammatic technique is given in Section {\bf
2}. By comparison with $N\times N$ case a diagrammatic proof of
semicircle density of states is found. We also develop partial
summation of infinite subseries of topologically trivial tree-type
diagrams. In Section {\bf 3} the ideology of double scaling limit is
used to study the edge of spectrum for random band matrices. The edge
of spectrum for lattice model (\ref{lattice}) is considered in Section
{\bf 4}. Surprisingly the critical behavior at the edge for lattices
with random hopping coincides with the critical behavior of the
string-theory inspired model considered in \cite{Jan}. In Section {\bf
5} the two-point correlation function $\overline{ \rho(E_1)
\rho(E_2)_c}$ is calculated. More precisely we found the so called
smoothed correlation function in the large $b$ limit and the first
$\sim 1/b$ correction to it. Moreover together with the hypothesis
about the universality of spectrum fluctuations \cite{Aronov} this
correlation function allows to find the correct estimate of the
localization length for the eigenfunctions of band matrix. This
universality holds also for the $\sim 1/b$ correction to the
correlation function, even though the $\sim 1/b$ correction itself
turns out to be a subject of strong cancellations. Finally in the
Section {\bf 6} some quantities which have no analog for usual
$N\times N$ matrices are considered. These are the correlation
function of local density states $\overline{ \rho(E_1,i)
\rho(E_2,j)}_c$ and the usual density-density correlation function for
the banded matrices of finite size $N$.

\section{Diagrammatic technique}\label{sec:2}

It follows immediately from (\ref{band}), that only diagonal
terms survive in the averaged Green function
\bq\label{GE}
\overline{G_{i,j}} = G(E) \delta_{i,j} \,\,\, .
\ee
Let us expand $G(E)$ in a formal series
\bq\label{ser}
G = \fr{1}{N} \overline{ tr \fr{1}{E-H}} = \fr{1}{NE}
\sum_{n=0}^{\infty} \overline{ tr \left( \fr{H}{E}
\right)^n } \,\,\, . 
\ee
Now all we need is using the Wick theorem and the second moment
(\ref{band}) to calculate all the average values of the trace of
product of $H$ - matrices. In the standard Feynman diagram technique
each $H^n$ corresponds to $n$-legs vertex and the averaging in
(\ref{ser}) reduces to counting of the number of possible ways of
contracting these legs to each other. However, like in $2d$-gravity it
is more convenient to draw the dual Feynman graphs. For dual diagrams
each $H_{ij}$ corresponds to a segment with numbers $i$ and $j$ at the
ends, while $H^n$ corresponds to $n$-vertex polygon with matrix
indices $i_1,i_2,\dots i_n$ assigned to the vertices (see fig. 1a). It
is also useful to draw the arrow on each segment showing the direction
from first to second index. Within this language the Wick
contractions in (\ref{ser}) correspond to the glueing of pairs of
segments. Our aim now is to calculate the number of ways in which the
edges of the polygon may be glued into some closed surface.

For Hermitian matrices ($\Theta=0$ in (\ref{band})) the
segments should be glued in the opposite direction thus
forming the oriented surface. For symmetric matrices
($\Theta=1$) the nonoriented surfaces are also allowed (e.g. the
M\"{o}bius band). 

\begin{figure}
\epsfysize=6.3 truecm
\centerline{\epsffile{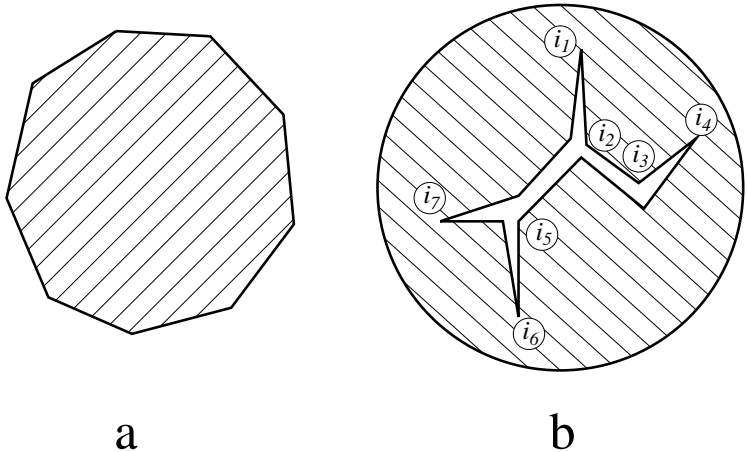}}
\caption{One plaquette a. and sphere b. glued from such plaquette}
\end{figure}

The example of simplest surface of spherical topology is
shown in fig. 1.b.  It is easy to verify that just
the spherical surfaces dominate in $G$ at large $b$. Only
in this case the sum over $n$ matrix indices for $H^n$
gives the factor $N b^{0}$ and thus (see
(\ref{band}),(\ref{defb})) 
\bq\label{Gsph}
G_{spherical} \sim b^0 \,\,\, .
\ee
Moreover, as may be seen from fig. 1b, the
summation over each index in spherical diagrams is
completely independent and results in a fixed factor $V^2$
for any choice of the function $F(i-j)$ (\ref{F}) and for
each of ensembles (\ref{band}),(\ref{lattice}),(\ref{NN}). In
particular this means (analogous observation was also done
in \cite{BZ}) that the Green function in the leading
approximation coincides for all ensembles
(\ref{band}),(\ref{lattice}),(\ref{NN}) 
\bq\label{G0}
G_0 = \fr{1}{2V^2} \bigl( E - \sqrt{E^2 -4V^2} \bigr) 
\ee
(a very clear proof of this formula for full $N\times N$
matrices see in \cite{Verb1,Pastur}).

Before considering the $\sim 1/b$ corrections let us carefully examine
equation (\ref{G0}). As we have said above, $G_0$ may be thought as
the exact sum of the part of the series (\ref{ser}) corresponding to
diagrams of spherical topology. It is seen from (\ref{G0}) that this
series is convergent only outside the circle on the complex $E$ plane
with a radius $|E|=2V$. The values of $G_0$ inside the circle ($|E| <
2V$) may be found via analytic continuation. This feature of the
series has two important physical consequences. Firstly, if one is
approaching the singular points $E=\pm 2V$ starting from large $E$ the
more and more complicated diagrams became important thus approaching
some kind of continuum limit. It may be not so obvious from the
equation (\ref{G0}) because for $G_0$ the series converges even at
$E\equiv \pm 2V$ but as we will see below the $\sim 1/b$ corrections
to $G$ are more singular and the summation of $(V/E)^n$ do is
saturated by the terms with very large $n$.

Even more troublesome turns out to be the calculation of correlation
functions. In this case one has to consider the Green functions close
to the borders of the cut ($E=E_0 \pm i0 \ , \ |E|<2V$) and very far
from the domain of convergency of the series(\ref{ser}). The price for
such unreliable procedure will be the severe cancellations of
different contributions to the $\sim 1/b$ corrections to correlation
function (see Section {\bf 5}).

As we have taken into account exactly in (\ref{G0}) all spherical
contributions, the $\sim 1/b$ corrections naturally turn out to be
determined by the diagrams of the more complicated topology. As we
have seen before (\ref{Gsph}),(\ref{G0}) the sum over spherical graphs
for full $N \times N$ matrices and for band matrices coincides up to
trivial replacement $N \rightarrow b$ because the summation over each
matrix index in the tree diagram of the kind of fig. 1b is independent
and results exactly in the trivial factor $V^2$.  This direct
correspondence does not holds for the diagrams of the more complicated
topology. However, each sum over matrix index contains effectively
$\sim b$ items and their magnitude corresponds to that for $N\times N$
matrices up to substitution $b \rightarrow N$. Thus all the business
with classification of the diagrams in the powers of the parameter
$1/b \ (1/N)$~, which was so productive for the full matrices, still
holds for the band matrices as well. In particular for the Hermitian
band matrices ($\Theta=0$ in (\ref{band})) one may use the well known
Eulers theorem to show that the corrections to (\ref{G0}) may be only
of the kind $\sim (1/b^2)^n$ with $n$ being the number of handles.

Up to now we have associated each Feynman diagram with some
surface. However, as it may be seen e.g.  from the fig.~1 because in
our problem we have in fact one large plaquette (or two for the
correlation functions) it is natural to consider only the border of
this plaquette which should be glued to some kind of branched polymer.
It is seen from the fig. 1b that the spherical surfaces are associated
with the tree--type polymers.  On the other hand the $\sim 1/b$
corrections to $G$ will be associated with selfintersections (the
closed loops) of the polymer. It seems very attractive to divide the
calculation of the $\sim 1/b$ corrections into two stages. The first
stage will consists in summation over the trees and at the second
stage one will take into account only the dressed selfintersected
diagrams. To this end it turns out to be useful to consider instead of
the Green function the logarithm
\bq\label{log}
L(E) = \fr{1}{N} \overline{ tr \ln \left( 1- \fr{H}{E} 
\right) } \ \ , \ \ G(E) = \fr{1}{E} + \fr{dL}{dE} \ .
\ee
The simple combinatorial calculation allows to replace the
perturbative series for $L(E)$ by the sum over skeleton graphs, as it
is demonstrated in the fig. 2
\begin{eqnarray}\label{scelet}
L(E) = -\fr{1}{N} \overline{\sum_{n=1} 
\fr{1}{n} tr \left( \fr{H}{E} 
\right)^n} = -\fr{1}{N} \overline{ 
\sum_{\begin{array}{c} {\scriptstyle no \ contractions} \\
                 {\scriptstyle of \ neighbours} \end{array}}
\fr{1}{p} tr \left( G_0 H 
\right)^p}
\end{eqnarray} 
Here the contractions between the nearest neighbours in the last sum
are forbidden because they have been taken into account exactly in
$G_0$-s. More precisely for the real symmetric matrices ($\Theta =1$
in (\ref{band})) the contractions between neighbours still are allowed
but only due to the second term ($\sim \Theta$) in the r.h.s. of
(\ref{band}).

\begin{figure}
\epsfxsize=11.9 truecm
\epsfysize=9.8 truecm
\centerline{\epsffile{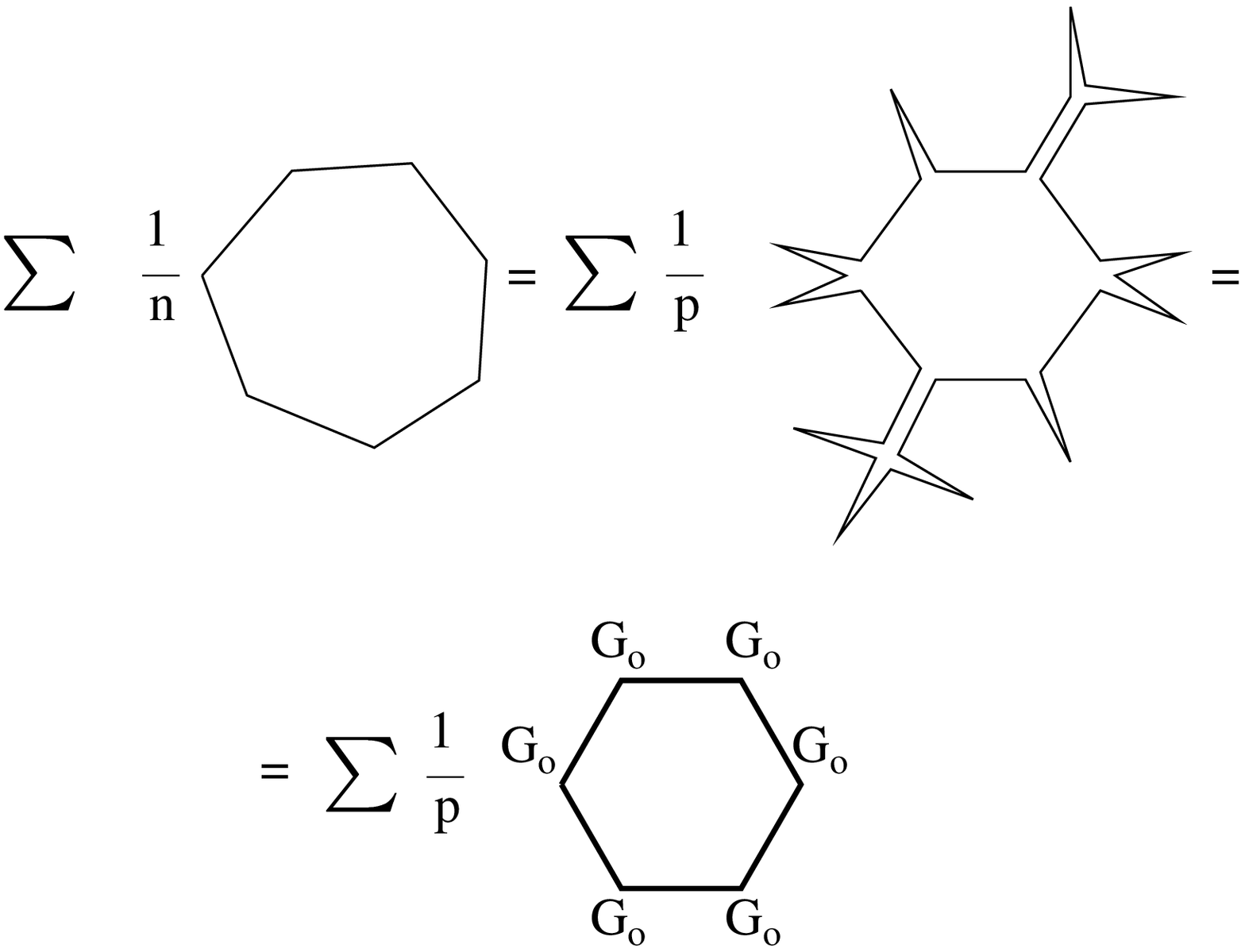}}
\caption{ Reduction of $\overline{\ln(1-E/H)}$ to skeleton 
diagram.}
\end{figure}

\section{The edge of spectrum}\label{sec:3}

Now at last we are able to consider the edge behavior of 
the $\sim 1/b$ corrections to the Green function $G_0$ (\ref{G0}).
To this end in particular we have to take into account the long 
chains of glued dressed links of the kind of (\ref{scelet}) (or
fig.~2). Consider the simplest two-link chain which is shown in
the fig.~3:
\begin{eqnarray}\label{chain}
\Psi_n(i_n -i_0 )&=& \sum_{i_1 i_2 \dots i_{n-1}}
(\overline{H_{i_0 i_1} H_{i_1 i_0}}) (\overline{H_{i_1 i_2}
 H_{i_2 i_1}})\dots (\overline{H_{i_{n-1} i_n}  H_{i_n
i_{n-1}}}) \\ &=& 
V^{2n} \fr{1}{b\sqrt{2\pi n}} \exp \left\{ - \fr{1}{2nb^2} (i_0
-i_n )^2 \right\} \ \ . \nonumber
\end{eqnarray}
Here in order to calculate $\Psi_n$ we have used the specific form of
the function $F(i-j)$ (\ref{F}). The various methods may be used
in order to prove (\ref{chain}). For example one may use the
mathematical induction method. While deriving the equation
(\ref{chain}) we have replaced all the summations over the
intermediate indices $i_1, i_2, \, ... \, , i_{n-1}$ by the
integrations. The accuracy of such procedure $\delta \Psi
\sim e^{-b}$ for any smooth function $F(i-j)$ still allows one 
to consider the $\sim (1/b)^n$ corrections for any $n<b$~.

\begin{figure}
\epsfysize=2.8 truecm
\centerline{\epsffile{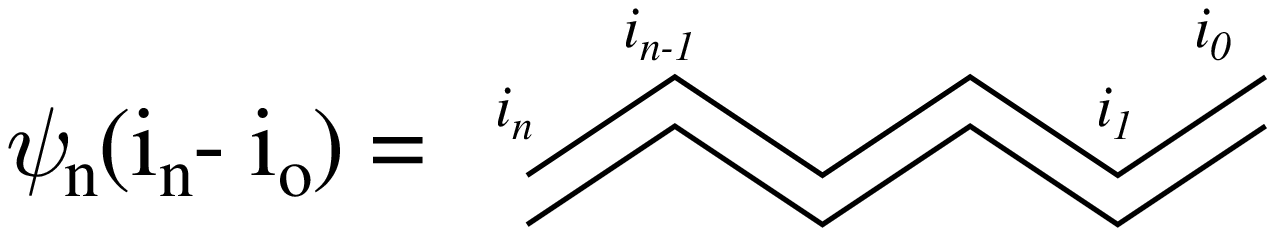}}
\caption{Double-line chain calculated in (13).}
\end{figure}

The $\Psi_n$ evidently satisfies the sum rule
\bq\label{rule}
\sum_{i_n} \Psi_n (i_n-i_0) \equiv 
 V^{2n} \ \ ,
\ee
which also may be used in order to find the normalization of
(\ref{chain}).

The sum rule (\ref{rule}) holds exactly for any choice of the function
$F(|i-j|)$ while the equation (\ref{chain}) is model
dependent. However, as it will be shown now, for large $n$ the
equation (\ref{chain}) is also universal. Consider the recursion
formula for large $n$ (and arbitrary $F$)
\bq\label{recur}
\Psi_{n+1}(i) = \sum_j F(|i-j|) \Psi_n(j) =
V^2\Psi_n(i) +\fr{V^2 b^2}{2}\fr{d^2}{di^2}\Psi_n(i)  
 + \, ... \ \ \ .
\ee
This equation is the discrete (in time) analog of
the heat conductivity equation. Together with the initial 
condition
\bq\label{initial}
\Psi_0(i) \sim \delta(i) \ \ ,
\ee
and the sum rule (\ref{rule}) equation (\ref{recur}) allows to
reproduce the formula (\ref{chain}) for the chain. In fact this
is the reason for considering the function $F(|i-j|)$ of the
form (\ref{F}) as the most universal one. Moreover, even if one
starts with some irregular function $F$ (which may seem to be
crucial because only for smooth $F(x)$ the summation may be
replaced by integration with the accuracy $\sim e^{-b}$), 
taking into account of a long chains
($n\gg 1$) effectively smooth it out.

Up to now we have considered only the band matrices. However the 
formula (\ref{chain}) may be easily generalized for the random-bond
lattice case (\ref{lattice})
\bq\label{latgen}
\Psi_n(\vec{i}\, )= \fr{V^{2n}}{b(2\pi n)^{d/2}} \exp \left\{ 
- \fr{1}{2nb^{2/d}} \vec{i}^{\, 2} \right\} \ \ .
\ee

\begin{figure}
\epsfxsize=5 truecm
\epsfysize=5 truecm
\centerline{\epsffile{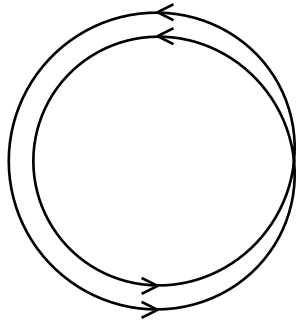}}
\caption{The M\"{o}bius band corresponding to the $\sim 1/b$
correction to Green function for real symmetric matrices.}
\end{figure}

At this point we have completed all the preliminary formalities and
are able to calculate the $\sim 1/b$ correction to the Green
function. Let us consider the ensemble of real symmetric matrices(GOE)
for which the correction of the first order $\sim 1/b$ exists. The
corresponding skeleton Feynman graph is shown in the fig.~4.  It is
seen from the figure that the collinear links should be glued in the
$(1/b)^1$ order and thus this diagram is forbidden for the
Hermitian(GUE) matrices. In terms of surfaces fig.~4 corresponds to
the M\"{o}bius band. Combining together (\ref{log}), (\ref{scelet})
and (\ref{chain}) one finds the correction
\bq\label{corr}
L_1 = 
\fr{1}{N} \overline{\delta \ln \left( 1-\fr{H}{E} \right)}
= -\fr{1}{N} \sum_{p=1}^{\infty} \fr{1}{2p} G_0^{2p} \,
\overline{tr H^{2p}}_{skeleton} =
-\fr{1}{b} \sum_{p=1}^{\infty} \fr{1}{2p} (G_0V)^{2p} 
\fr{1}{(2\pi p)^{d/2}} \ ,
\ee     
or for the Green function 
\bq\label{dgreen}
G_1 = \delta G= 
-\fr{1}{b(2\pi )^{d/2}} \fr{G_0'}{G_0} 
\sum_{p=1}^{\infty} \fr{(G_0V)^{2p}}{p^{d/2}} \ \ .
\ee     
We give the result at once for arbitrary dimensionality $0
\le d < \infty$. For band matrices one has to choose $d=1$.

The only important factor which is responsible for the difference
between usual $N\times N$ matrices ($d=0$), band matrices and, random
hopping Hamiltonians in (\ref{dgreen}) is the $1/p^{d/2}$ in the
sum. Technically this factor comes from the double chain of matrix
elements (\ref{chain},\ref{latgen}). One may consider the length of
the M\"{o}bius band (fig. 4) $p$ as the discrete time in the diffusion
equation (\ref{recur}). Then the $1/p^{d/2}$ will simply correspond to
the time dependance of the return probability for classical diffusive
particle at the time $p$ for different dimensions.

We are mostly interested in the singularities of $G$ at $G_0
V\rightarrow 1$ (or $E\rightarrow \pm 2V$, see (\ref{G0})).  It
is seen from (\ref{corr}) that $L_1$ has a finite limit at $G_0 V
=1$ for any $d > 0$. On the other hand the $G_1$ (\ref{dgreen})
is convergent and the diagram of fig.~4 approaches the continuum
limit, at least for $d=1$ and $d=2$ (for $d>2$ the $G_1$ is also
singular, but mostly due to $G_0'$ without any continuum limit).

For the random band matrix case ($d=1$) it is easy to found from
(\ref{dgreen}), (\ref{G0}) that close to singular point
\bq\label{expan}
G(E \rightarrow 2V) = \fr{1}{V} -\fr{1}{V}\sqrt{\fr{E-2V}{V}}
+ \fr{1}{b} \, \fr{1}{4 V} \left( \fr{V}{E-2V} 
\right)^{3/4} + \fr{1}{b^2} \fr{const}{V} \left( \fr{V}{E-2V} 
\right)^2 + \, ... \ \ .
\ee
Here the last term is the order of magnitude estimate of the 
$\sim 1/b^2$ contribution. The Feynman diagrams corresponding
to this contribution are shown in the fig. 5 (the explicit 
calculation of one of them will be presented below). 
\begin{figure}
\epsfxsize=10.5 truecm
\epsfysize=7 truecm
\centerline{\epsffile{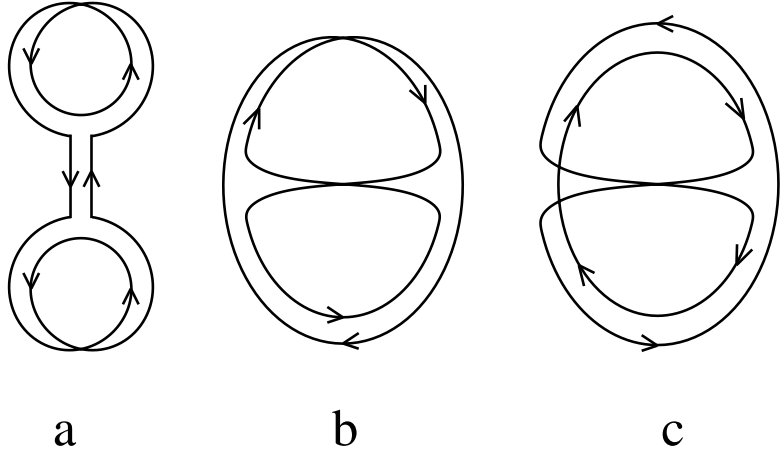}}
\caption{The three possible second order $\sim 1/b^2$ 
diagrams for ensemble of real symmetric (GOE) matrices,
a. -- dumb-bell 
shaped diagram, c. -- torus}
\end{figure}
The simple
counting of the power of convergency for the higher order
diagrams shows that close to singularity the Green function
should be described by some scaling function
\bq\label{scaling}
G = \fr{1}{V} +\fr{1}{V}\fr{1}{b^{2/5}} \, \Phi\left( b^{4/5}
\bigg( \fr{E-2V}{V} \bigg) \right) \ \ .
\ee

So, one may conclude that the singularity at the edge $E=\pm 2V$ of
the perturbative Green function (\ref{G0}) should be smoothed out at
the distances $\Delta E \sim V b^{-4/5}$ from the singular points. For
example it is easy to estimate the number of energy levels falling
into this region. The same estimate evidently holds for the number of
the levels outside the circle $|E|>2V$
\bq\label{DeltaN}
\Delta N \sim N \Delta E^{3/2} \sim \fr{N}{b^{6/5}}
\ \ .
\ee
This estimate is of particular interest because the density of
eigenvalues outside of the circle is purely nonperturbative and
could not be found in any finite order over $1/b$.  It is to be
noted that for usual $N\times N$ matrices (both GOE and GUE) 
$\Delta E \sim N^{-2/3}$ and $\Delta N \sim 1$.

As we have told in the introduction, random band matrix may be
considered as a Hamiltonian for particle on $1d$ lattice with random
hopping. Eigenvectors for such Hamiltonian should naturally have some
finite localization length. This localization length for the band
matrix ensembles (\ref{band}) was found in the papers of Fyodorov and
Mirlin \cite{Mirlin} within the super-symmetry method (the calculation
of $l_{loc}$ in our approach will be given in Section {\bf 5})
\bq\label{Fyod}
l_{loc} \sim b^2 \left( 1-\fr{E^2}{4V^2}\right) 
\ \ , \ \ |E|<2V \ \ .
\ee
This result in fact was found only in the leading order in $1/b$ and
should be changed at the edge of spectrum. One may combine
our result (\ref{scaling}) with (\ref{Fyod}) in order to estimate the
localization length for nonperturbative states living outside of the
main band (circle)
\bq\label{lnp}
l_{loc}^{np} \sim b^2 \Delta E \sim b^{6/5} \ \ .
\ee
In particular $l_{loc}^{np} \Delta N \sim N$ which may mean that
nonperturbative states spatially do not overlap.

In the recent paper \cite{Felix} the distribution of Lyapunov
exponents for random band matrices was studied numerically at the very
edge of perturbative part of spectrum $ |E| = 2V$. The authors of
ref. \cite{Felix} have considered the band matrices with step form of
the band $F(|i-j|>b) \equiv 0$ for which there exist exactly $2b+1$
Lyapunov exponents. Their result reads
\bq\label{Lyap}
\lambda_i \sim b^{-4/3} i^{1/3} \ \ \ .
\ee
Here $i=1,2,3 ...$ , and $\lambda_1$ is the smallest decrement of the
Lyapunov exponent. The corresponding solution grows like $\psi_n^{(i)}
\sim \exp(n\lambda_i)$. Naively one may expect that the smallest
Lyapunov exponent directly gives the localization length
\bq\label{Lyaploc}
\lambda_1^{-1} \approx l_{loc} \ \ \ .
\ee
However, this simple guess puts in conflict the two results
(\ref{lnp}) and (\ref{Lyap}). On the other hand, the numerical
accuracy of \cite{Felix} allows one to be sure in (\ref{Lyap}) for
sufficiently large $i$, namely $1\ll i\ll b$. The accurate result for
smallest Lyapunov exponent still may differ from (\ref{Lyap}).

Moreover, the direct correspondence between the localization length
and first Lyapunov exponent naturally take place in the central part
of spectrum $|E|<2V$ but at the edge $|E|=2V$ formula (\ref{Lyaploc})
is not necessary correct. It seems rather probable that the
eigenstates outside the circle $|E|\ge 2V$ appears due to some very
rare fluctuations of our random banded Hamiltonian
$H_{ij}$. Furthermore, we have no idea how even to estimate the
density of such rare fluctuations. If the number of this fluctuations
is small compared to $N/b^{6/5}$ the eigenstates with $|E|>2V$ will
not be distributed homogeneously, but will be concentrated in rather
rare and dense bunches. On the other hand, the first Lyapunov exponent
$\lambda_1$ evidently comes from the whole range of variation of
vector index. Thus the localization length at $|E|=2V$ should not be
necessary of the same order of magnitude with the first Lyapunov
exponent.  Unfortunately, if the states with $|E|>2V$ do are
concentrated together one should most naturally expect that
$\lambda_1^{-1} \gg l_{loc}^{np}$. Thus, the discrepancy between
(\ref{lnp}) and (\ref{Lyap}) turns out to be even severe in this case.

\begin{figure}
\epsfxsize=10 truecm
\epsfysize=6.5 truecm
\epsfxsize=8 truecm
\epsfysize=5.2 truecm
\centerline{\epsffile{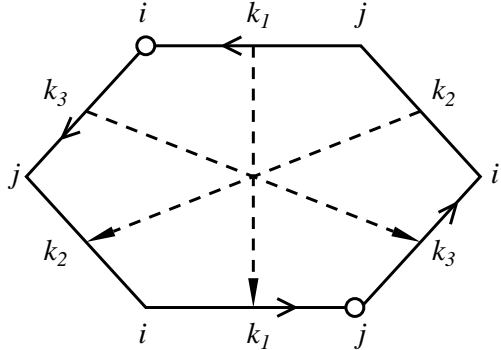}}
\caption{Gluening of dressed hexagon for $\sim 1/b^2$ correction 
for Hermitian matrices (torus).}
\end{figure}

In fact in (\ref{expan}), (\ref{scaling}) the estimate for higher
order corrections was declared without any proof.  However, both for
the more rigorous proof of (\ref{scaling}) and for future calculations
it is useful to calculate explicitly at least one nontrivial (beyond
the one loop) diagram. Therefore we would like to found now the
leading $\sim 1/b^2$ correction for the GUE band matrices. In this
case the only diagram of fig. 5c survive which corresponds to torus in
the surface language.  The fig. 6 shows how one have to glue the
dressed hexagon (see (\ref{scelet})) in order to built this
diagram. Indices $k_1,k_2,k_3$ on the figure are the lengths of the
double-link chains (\ref{chain}), while $i$ and $j$ stands for the
matrix indices corresponding to the ends of these segments. It is to
be noted that both indices $i$ and $j$ appear three times in the
diagram, which in particular leads to loss of two powers of large $b$
for this diagram compared to the tree one (fig. 1b).  More precisely,
as it may be seen also from the diagram fig. 1b not the multiple
indices but the closed loops are the direct source of the $1/b$~. One
of the simplest ways to find the order of the diagram in the $(1/b)^n$
classification is to look for the number of links which are enough to
break in order to get the tree-kind simply connected diagram. Now one
may combine (\ref{scelet}) and (\ref{chain}) in order to found the
contribution of the fig. 5c
\begin{eqnarray}\label{L2H0}
 L_2^H &=& -\fr{1}{N} \sum_{k_1,k_2,k_3>0} 
\fr{1}{2(k_1+k_2+k_3)} (k_1+k_2+k_3) \fr{1}{3}
(G_0)^{2k_1+2k_2+2k_3} \\
&\times& \sum_{i,j} 
\Psi_{k_1}(i-j)\Psi_{k_2}(i-j)\Psi_{k_3}(i-j)
\ \ . \nonumber
\end{eqnarray}
Here $(k_1+k_2+k_3)^{-1}$ comes from $1/p$ in (\ref{scelet}) and the
combinatorial factor $(k_1+k_2+k_3)^{+1}$ takes into account the
number of positions available, say, for the left end $i$ of the upper
segment $k_1$ on the fig. 6 and the right end $j$ of the same down
segment (these points are shown by the two small circles on the
figure). Because these two circles $i$ and $j$ are equivalent the
combinatorial factor is only $\sum k_i$~, not the $2\sum
k_i$~. Finally, after one has taken into account by this $\sum k_i$
the freedom in definition of the starting point on the circle the sets
$(k_1,k_2,k_3)$, $(k_2,k_3,k_1)$ and $(k_3,k_1,k_2)$ became
indistinguishable which is taken into account by the factor $1/3$ in
(\ref{L2H0}). The simple substitution of (\ref{chain}) into
(\ref{L2H0}) leads to
\begin{eqnarray}\label{L2H}
&\,& L_2^H = \\ 
&\,& -\fr{1}{6b^2} \sum_{k_1,k_2,k_3>0} \fr{(
G_0V)^{2k_1+2k_2+2k_3}}{2\pi \sqrt{k_1k_2k_3}} \sum_i
\fr{1}{b\sqrt{2\pi}} 
\exp\left\{ -\fr{1}{2 b^2}\bigg( \fr{1}{k_1}+\fr{1}{k_2}
+\fr{1}{k_3}\bigg) (i-j)^2 \right\} \nonumber \\
&\,& \ \ \ = -\fr{1}{12\pi b^2} \sum_{k_1,k_2,k_3>0} \fr{(
G_0V)^{2k_1+2k_2+2k_3}}{\sqrt{k_1k_2+k_1k_3+k_2k_3}} 
\ \ . \nonumber
\end{eqnarray}

Thus the Green function close to the edge of spectrum for the
GUE band matrices reads
\begin{eqnarray}\label{GH}
G^H &=& \fr{1}{V} -\fr{1}{V}\sqrt{\fr{E-2V}{V}}
+ \fr{1}{b^2} \fr{I}{V} \left( \fr{V}{E-2V} 
\right)^2 + \, ... \, = \nonumber \\
&=& \fr{1}{V} +\fr{1}{V}\fr{1}{b^{2/5}} \, 
\tilde\Phi\left( b^{4/5}
\bigg( \fr{E-2V}{V} \bigg) \right) \ \ ,
\end{eqnarray}
where
\bq\label{Integral}
I= \int_{x,y,z>0} \fr{\delta (1-x-y-z)}{48\pi}
\fr{dxdydz}{\sqrt{xy+xz+yz}} \ \ .
\ee
In order to be sure that we have done nothing wrong with the
combinatorics one may easily repeats the calculation
(\ref{L2H0}-\ref{GH}) for usual $N\times N$ matrices. Some new
(unknown) scaling function $\tilde \Phi$ appeared in (\ref{GH}).
We have used the same argument for $\tilde \Phi$ as for $\Phi$
(\ref{scaling}) but the asymptotic series for $\Phi(x)$ is in
powers of $x^{-5/4}$~, while for $\tilde
\Phi(x)$ in powers of $x^{-5/2}$~.

\section{The edge of spectrum for lattices}\label{sec:4}

Even more puzzling turns out to be the edge of spectrum behavior for
the lattice ensembles (\ref{lattice}). On the one hand, the naive
power counting for diagrams like, e.g., the diagram of fig.~5c which
we have calculated in (\ref{GH}) gives
\bq\label{wrongsc}
G = \fr{1}{V} +\fr{1}{V}\fr{1}{b^{\fr{2}{6-d}}} \, \Psi\left(
b^{\fr{4}{6-d}} 
\bigg( \fr{E-2V}{V} \bigg) \right) \ \ .
\ee
On the other hand, it is easy to find the first order ($ \sim
1/b$) correction to the Green function at the edge from
(\ref{dgreen}) 
\begin{eqnarray}\label{G1l}
G_1=\fr{1}{8\pi bV} \sqrt{\fr{V}{E-2V}} 
\ln\left(\fr{V}{E-2V}\right) \ \ &,& \ \ d=2 \ \ \ , \\
G_1=\fr{1}{2bV} \sqrt{\fr{V}{E-2V}} 
\sum_{p=1}^{\infty} \fr{1}{(2\pi p)^{d/2}} \ \ &,& \ \ d>2 \ \ \  .
\nonumber
\end{eqnarray}

It is seen immediately that at least for $d>2$ equations (\ref{G1l})
and (\ref{wrongsc}) disagree. In fact the solution to this paradox was
found a few years ago in a paper of Ambj{\o}rn et al. \cite{Jan} there
the toy model for string in $d$-dimensions has been considered. The
authors of ref. \cite{Jan} have restricted the class of triangulations
for string embedded in $d$-dimensions to those having minimal
cross-section.  While doing so they have obtained effectively the
theory of $d$-dimensional branched polymer. As we have mentioned
above, the summation over dual Feynman diagrams for our band
matrices--lattices also reduces to summation over some branched
polymers. In the lattice case indices assigned to the ends of each
link expands over $d$-dimensional (although discretized) Euclidean
space while the factor $F(\vec{i} -\vec{j} )$ (\ref{lattice})
regulates the spatial size of the link just like in the model of
\cite{Jan}. The branched polymers were also considered many times
within the random vector--matrix models approach (see
e.g. \cite{polymers}). However, only the critical exponents for our
model of branched polymers (or it is better to say ``branched tapes''
as it is seen from the figs.~4,5,7) should coincide with those for
another models. The scaling function itself may be different.

\begin{figure}
\epsfxsize=12 truecm
\epsfysize=8 truecm
\centerline{\epsffile{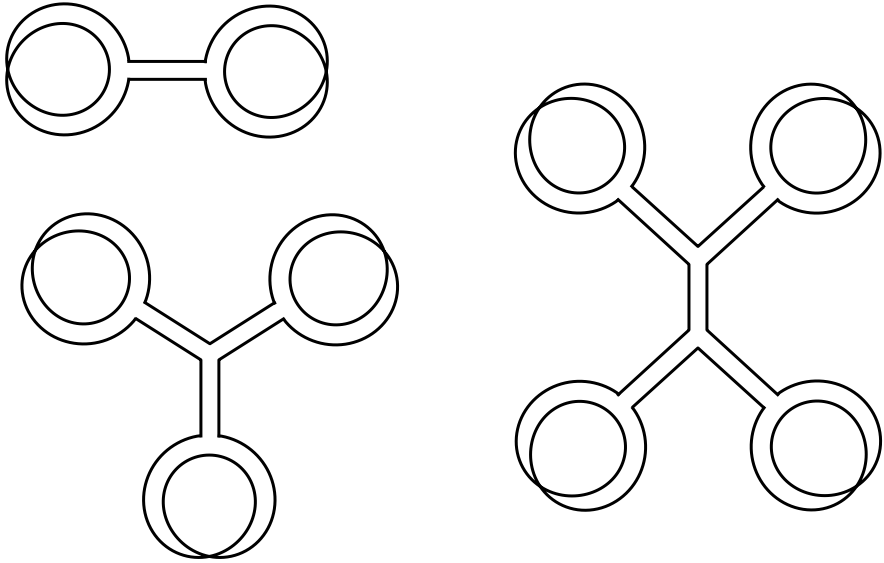}}
\caption{Tadpoles or new trees for lattice 
models.}
\end{figure}

In order to solve the contradiction between (\ref{wrongsc}) and
(\ref{G1l}) it is enough to observe that for $d\ge 2$ some of the
diagrams for the branched polymer are much more singular than others
with the same topology. These are the so called tadpole diagrams shown
in fig.~7. Moreover, each of the diagrams of fig.~7 behaves
effectively like some tree diagram (see fig. 1b) up to some trivial
factor associated with the ends of the tree. Therefore, while the true
complicated diagrams (say of fig. 5b,c) became less and less singular
in higher dimensions in accordance with (\ref{wrongsc}) the tadpole
diagrams for $d>2$ have all the same singularity.  On the other hand,
one may try to sum up exactly this subseries of rather simple
diagrams, as it was done in \cite{Jan}.

\begin{figure}
\epsfxsize=12.6 truecm
\epsfysize=6.3 truecm
\centerline{\epsffile{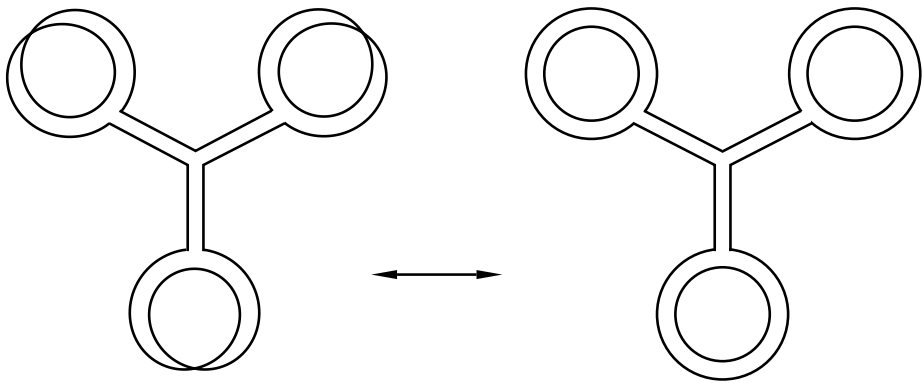}}
\caption{The reduction of tadpole diagram to corresponding
correlation function.}
\end{figure}

The procedure of calculation of the tadpole contribution is
illustrated by the fig.~8. Let us replace each $n$-tadpole
diagram by some correlation function of our Green function and
$(tr H)^n$ calculated in the spherical approximation. Explicit
formula corresponding to such procedure is as follows
\bq\label{tadsum}
G_{tadpole} = \fr{1}{N} \overline{\sum_n \left( tr \fr{1}{E-H}
\fr{(\kappa \, tr H)^n}{n!} \right)_c } \ \ \ ,
\ee
where subscript $c$ stands for connected diagrams. The coefficient
$\kappa$ may be found for example from (\ref{G1l}). The factor $1/n!$
in (\ref{tadsum}) accounts for the permutations of various $trH$. Due
to this $1/n!$ the sum over $n$ results in the trivial exponentiation
of $\kappa\, tr H$.  In order to find $G_{tadpole}$ it is useful to
write explicitly the integration over $H$
\bq\label{tadint}
G_{tadpole} = \fr{1}{Z} \int DH \fr{1}{N} tr \fr{1}{E-H} \exp
\big\{ -\sum_{i,j} H_{ij} H_{ji} M(i-j) +\kappa \, tr H
\big\} \ \ \ .
\ee
Here $M(i-j) \sim F(i-j)^{-1}$ and $Z$ is the same integral without 
$\fr{1}{N} tr \fr{1}{E-H}$ but with $\kappa \, tr H$
included in the argument of the exponent. By such choice of $Z$
one gets rid of disconnected diagrams. 

At least in the leading (spherical) approximation formulas
(\ref{tadsum}) and (\ref{tadint}) give the same Green function
at the edge. On the other hand, the Green function
(\ref{tadint}) reduces to the zero order one (\ref{G0}) by
trivial substitution
\begin{eqnarray}\label{shift}
 H_{ij} &\rightarrow& H_{ij} + const \, \delta_{ij} \ \ \ \ , \\
 E &\rightarrow& E + const \ \ \ \ . \nonumber
\end{eqnarray}
After comparison with (\ref{G1l}) one finds
\begin{eqnarray}\label{Gtadpole}
G=\fr{1}{V} -\fr{1}{V} \sqrt{\fr{E-2V}{V} - \fr{1}{4\pi b} 
\ln\left(\fr{V}{E-2V}\right)} \ \ &,& \ \ d=2 \ \ \ , \\
G=\fr{1}{V}-\fr{1}{V} \sqrt{\fr{E-2V}{V}
- \fr{1}{b} \sum_{p=1}^{\infty} \fr{1}{(2\pi p)^{d/2}} }\ \ &,& \ \ d>2
\ \ \  . 
\nonumber
\end{eqnarray}
For $d>2$ this result is almost trivial. Taking into account of the
most singular series of corrections result in a simple shift of the
edge of the cut. For $d=2$ taking into account of the tadpoles results
not only in the shift of the edge by $\Delta E\sim \ln(b)/b$ but also
in some nontrivial change of Green function. The singularity in
(\ref{Gtadpole}) for $d=2$ will be smoothed out at $|E-2V
-\fr{V}{2b}\ln(b)| \sim \fr{V}{b}$ by some unknown scaling function.

For GUE matrices(lattices) the simple tadpoles consisting of small
M\"{o}bius band are forbidden. However one may arrange slightly more
complicated $\sim 1/b^2$ tadpole shown on the fig.~9 which is allowed
also for oriented surfaces. As a result for the GUE lattices with
random bonds the critical dimension is $d=4$ instead of $d=2$.

\begin{figure}
\epsfxsize=6.3 truecm
\epsfysize=4.2 truecm
\centerline{\epsffile{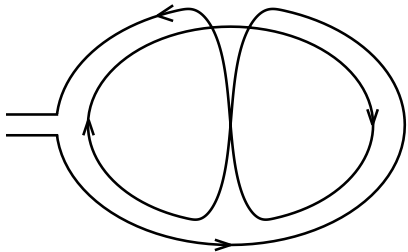}}
\caption{The $\sim 1/b^2$ tadpole for Hermitian(GUE) matrices.}
\end{figure}

\section{Correlation function}\label{sec:5}

Physically may be the most interesting quantity which would be
calculated with the random matrix ensembles is the two point
correlation function of density of states at very close energies. It
is generally believed that just such local quantities most adequately
reproduce the measurable features of complicated quantum systems.

Unfortunately the perturbative procedure, which we are only able to
perform, has a serious drawback in the case of correlation
functions. As it was said in the Section {\bf 2}, the perturbative
series in $1/E$ are convergent only outside the circle $|E|>2V$ while
the series in $1/b$ turns out to be the asymptotic series. Besides the
perturbation theory there may exist some nonperturbative contributions
say of the form $\sim \exp(-b(E-2V)^\gamma)$ with some $\gamma \sim 1$
. However, after analytic continuation to the border of the cut, which
goes from $E=-2V$ to $E=2V$ , these corrections may be (and do are at
least for the usual $N\times N$ matrix ensembles) converted into some
oscillating functions like $\sim \sin (b^2 \Delta E)$~. Certainly in
our perturbative result all these oscillating terms (if they are) will
be smoothed out.  For example for the usual GUE ensemble one gets
instead of the exact result $K(\Delta E)=\sin^2(N\Delta E)/\Delta E^2$
only $K=0.5/\Delta E^2$ . Due to this principal limitation the
quantities which we would calculate are called the smoothed
correlators.

The intriguing feature of our diagrammatic series which was considered
in the previous sections is that the dressed Feynman diagrams at the
edge of spectrum approach some kind of continuum limit. Technically it
happens because the value of $(VG_0(E))^2$ (\ref{G0}) which was our
actual expansion parameter became equal to $(VG_0(2V))^2=1$ at the
border. Now, because we are going to work with the energies inside the
band $|E_{1,2}|<2V$ there seems to be no room for the continuum
limit. However, let $E_1$ approach the upper border and $E_2$ approach
the lower border of the cut . Then it follows immediately from
(\ref{G0}) that
\begin{eqnarray}\label{GGG}
V G_0(E_1)&=&e^{i\phi-\lambda/2} \ \ \ , \\
V G_0(E_2)&=&e^{-i\phi-\lambda/2} \ \ \ , \nonumber
\end{eqnarray}
where small $\lambda $ is given by
\begin{eqnarray}\label{LamLL}
\lambda = \fr{E_1 -E_2}{i\sqrt{4V^2-E^2}}
\ \ \ , \ \ \ 
Re\lambda &>& 0 \ \ \ \ . 
\end{eqnarray}
Thus again at least those subseries of diagrams which will have
as the expansion parameter the combination $V^2 G_0(E_1)G_0(E_2)
=e^{-\lambda}$ may approach the continuum limit (another less
direct example of the ``effective'' continuum limit will be
considered in the following subsection). 

In order to find the correlation function of two Green functions
it is natural again to consider the logarithms 
\begin{eqnarray}\label{LLL}
&\,&
\overline{Tr \ln\left( 1-\fr{H}{E_1} \right) 
          Tr \ln\left( 1-\fr{H}{E_2} \right)}_{c} \equiv
\nonumber \\
&\equiv&\sum_{p,p'} 
\overline{\fr{1}{p'} Tr\bigg\{ (G_0(E_1)H )^{p'}
          \bigg\}_{skelet}  
          \fr{1}{p} Tr\bigg\{ (G_0 (E_2)H )^p 
          \bigg\}_{skelet}} = \\ 
&=&2\fr{N}{b(2\pi)^{d/2}}\sum_p \fr{p}{p^2} \fr{1}{p^{d/2}}
\big( V^2 G_0(E_1) G_0(E_2)\big)^p \ \ \ . \nonumber
\end{eqnarray}
Here we have expanded each logarithm in the sum of closed skeleton
chains like in eq. (\ref{scelet}). Also here and below subscript $c$
means ``connected''. A factor $2$ in front of the last sum accounts
for two allowed directions of glued chains (Cooperon and diffuson in
solid state physics).  Finally, one combinatorial factor $p$ in the
last sum accounts for the number of different ways to contract two
skeleton rings of the length $p$. We see that just the $e^{-\lambda}$
(\ref{LamLL}) turns out to be the expansion parameter in
(\ref{LLL}). 

By simple differentiation of (\ref{LLL}) one finds the
correlation function at very close energies $E_1-E_2 \rightarrow 0$
\begin{eqnarray}\label{Altsh}
\overline{Tr \fr{1}{E_1-H}  
          Tr \fr{1}{E_2-H}}_{c} &=&
\fr{2N}{b(4V^2-E^2)} \sum_{p=1}^{\infty} 
\fr{p^{1-d/2}}{(2\pi)^{d/2}} e^{-\lambda
p} \sim \\
&\sim& \left\{ \begin{array}{ll}   
               \left(\fr{i}{E_1 -E_2}\right)^{2-d/2} \ \ ; \ \
    d<4 \\ \ \\
               \ln\left(\fr{iV}{E_1 -E_2}\right) \ \ ; \ \ d=4
               \end{array} \right. 
                \ \ . \nonumber
\end{eqnarray}
This equation reproduces the known result of Al'tshuler and Shklovskii
\cite{Altshuler} for correlation of energy levels in small disordered
metallic samples.

The factor $1/p^{d/2}$ in (\ref{LLL},\ref{Altsh}) appears after
glueing of two $H^p$ into closed two-link chain
(\ref{chain},\ref{latgen}). As it was said also after
eq. (\ref{dgreen}) this factor $1/p^{d/2}$ works effectively as the
probability for diffusive particle to return to origin after time
$p$. In principle one may go even further in this analogy with
classical diffusion. By making the Fourier transform of the function
$\Psi_n(\vec{i}-\vec{j})$ (\ref{chain},\ref{latgen}) the equation
(\ref{Altsh}) may be written in the form of the integral of squared
Green function of the diffusion equation (again in agreement with
corresponding formulas from \cite{Altshuler}).

Now let us consider in more details the correlation function for band
matrices $(d=1)$. We would like to show how the information contained
in the smoothed correlation function (\ref{Altsh}) combined with the
simple hypothesis of universality of spectral correlations allows to
find the correct estimate of the localization length for eigen-vectors
of random band matrices.  Consider instead of Green function $G$ the
density of eigen-values $\rho(E)$ , which is simply the imaginary part
of $G$ (\ref{Green}). It is generally recognized, that the
fluctuations of $\rho(E)$ for all ensembles of full random $N\times N$
matrices are universal. This means that, being properly normalized,
the density of states -- density of states correlation function has
the form:
\bq\label{corrr} \fr{\overline{\rho(E_1) \rho(E_2)_{c}}}{
\overline{\rho(E_1)} \ \overline{\rho(E_2)} } =
K\left(\fr{dn}{dE}(E_1-E_2) \right) \ \ \ \ .  \ee Here the function
$K(x) \sim 1$ at $x \sim 1$ and $dn/dE$ is the averaged density of
eigenstates. Of course, both $E_1$ and $E_2$ in (\ref{corrr}) are real
(have reached the border of the cut).  The detailed form of $K(x)$ is
specific for the ensemble under consideration (e.g. GUE or GOE), but
for given ensemble $K$ is the universal function of energy interval
$\Delta E = E_1 -E_2$ measured in units of mean inter-level spacing
$dE/dN$ .

For the random band matrices one may expect the same universal
behavior of the correlation of density of state fluctuations as
(\ref{corrr}) only if all eigen-vectors are delocalized. On the other
hand it is easy to write down the natural extension of the formula
(\ref{corrr}) for systems with finite localization length $l$ :
\bq\label{corrl}
\fr{\overline{\rho(E_1) \rho(E_2)_{c}}}{
\overline{\rho(E_1)} \ \overline{\rho(E_2)} } =
\fr{l}{N} K\left(\fr{l}{N} \fr{dn}{dE}( E_1-E_2) 
\right) \ \ \ \ . 
\ee
By writing this formula we suppose that the fluctuations of
density of energy levels for random band matrices still are
universal if the energy difference $\Delta E = E_1 -E_2$ is
measured in the units of effective mean inter-level spacing  
\bq\label{Eeff}
\Delta E_{eff} \sim \fr{N}{l} \fr{dE}{dn} \ \ \ \ .
\ee
Roughly speaking $\Delta E_{eff}$ is the mean inter-level spacing for
a band matrix of a finite size $N\sim l$ (of course in
(\ref{corrr},\ref{corrl}) we suppose that at least $l \gg b$
). Moreover, if the modified universality do takes place and $l \gg b$
the formulas (\ref{corrl},\ref{Eeff}) will also account for the energy
dependence of the localization length. Of course the two universal
functions $K(x)$ in (\ref{corrr}) and (\ref{corrl}) are completely
different.

Our line of reasoning in fact follows the consideration of
\cite{Aronov} (see also \cite{Atland}) for density-density correlator
in disordered metal. Now we would like simply to convert the arguments
of authors of ref. \cite{Aronov} in order to estimate the localization
length. The averaged density of states for large $b$ may be easily
found from (\ref{G0})
\bq\label{rho0}
\overline{\rho(E)} = \fr{dN}{dE} =  \fr{1}{\pi} Im \,
\overline{tr G(E-i0)} = \fr{N}{2\pi V^2} \sqrt{4V^2 -E^2} \ \ .  
\ee
Now from (\ref{Altsh}) one finds the asymptotics
\bq\label{corra}
\fr{\overline{\rho(E_1) \rho(E_2)_{c}}}{
\overline{\rho(E_1)} \ \overline{\rho(E_2)} } =
\fr{-1}{Nb} \ \fr{V^4}{(4V^2 -E^2)^{5/4}}
\ \fr{1}{|E_1-E_2|^{3/2}} \ \ \ . 
\ee
Together with (\ref{corrr}) and (\ref{corrl}) this correlation
function allows to find:
\bq\label{loc}
l= \ b^2 \left( 1 -\fr{E^2}{4V^2} \right) 
\ee 
in accordance with the result of Fyodorov and Mirlin \cite{Mirlin}. We
have defined here the localization length (or effective localization
length) by choosing the overall normalization constant in (\ref{loc})
to be equal to one. Formally, universa\-li\-ty--based arguments allows
to find $l$ only up to some normalization constant $\sim 1$ , which on
the other hand, depends on the explicit definition one uses for the
localization length. For example, $l$ defined via the inverse
participation ratio \cite{Mirlin} or by the first Lyapunov exponent
may differ by some trivial factor.  Anyway, our estimate $l(E)$ seems
to be much less complicated than those of the supersymmetric approach
of \cite{Mirlin}.

\subsection{Corrections to the correlator}\label{sec:5.1}

In this subsection we would like to consider the $\sim 1/b$
corrections to the correlation function. As it was explained above, we
are able to calculate only the smoothed correlation functions. More
concretely we are able to consider the Green functions not too close
to the border of the cut $|Im E|\gg \Delta E_{eff} \sim V/b^2$
(\ref{Eeff},\ref{loc}).  Nevertheless, the uncertainty in the
correlation function due to smoothening of fast oscillations decrease
exponentially like $\sim \exp (-ImE/\Delta E_{eff})$. Therefore even
for smoothed correlation function one is able to consider the
corrections of any finite order in $1/b$.

The skeleton diagrams for $\sim 1/b$ correction to the correlation
function (\ref{Altsh},\ref{corra}) are shown in the fig.~10. At a
first stage one may easily estimate the power of singularity for each
diagram at small $\Delta E$. This singularities are associated with
the number of summations over the length of tape glued from the
segments originated from two different logarithms (\ref{LLL}) $\
\ln(1-H/E_1)$ and $\ \ln(1-H/E_2)$. Thus compared to the zeroth order
result (\ref{Altsh},\ref{corra}) the diagrams of fig.~10a. are of the
relative amplitude $\sim b^{-1} \Delta E^{-2+d/2}$ while the diagrams
10b. are of the relative amplitude $\sim b^{-1} \Delta E^{-1}$~.
\begin{figure}
\epsfysize=8.4 truecm
\centerline{\epsffile{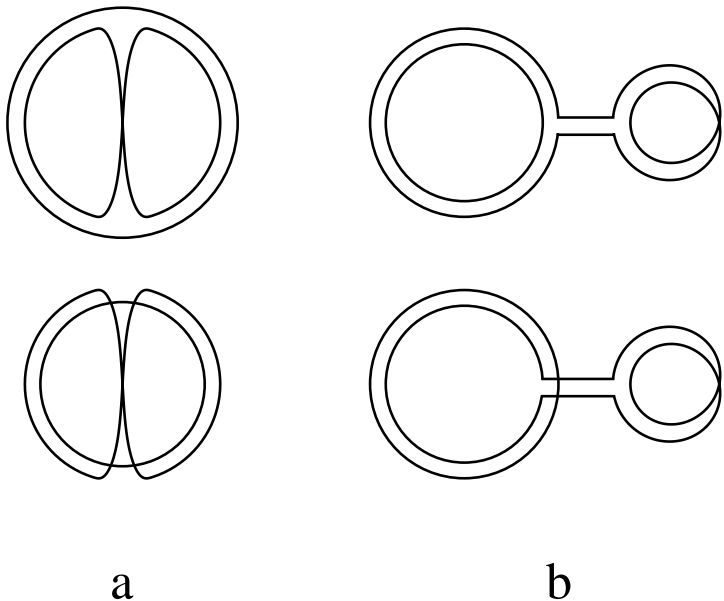}}
\caption{The $\sim 1/b$ corrections to correlation
function. a. corrections singular like $1/\Delta E^3$ for $d=1$, b.
corrections singular like $1/\Delta E^{5/2}$ for $d=1$.}
\end{figure}

On the other hand, the hypothesis of universality (\ref{corrl}) which
allows us to find so successfully the localization length (\ref{loc})
leads to a strong restriction on a possible form of corrections to the
correlation function. As we have considered in the Section 2, the
spherical result for the correlation function (\ref{corra}) is the
exact result in the large $b$ limit. Therefore the corrections to
(\ref{corra}) should be of the relative order $\sim 1/b$. In its turn
due to the universality (\ref{corrl}) $b$ may appear in the result
only in the combination $(E_1-E_2)/\Delta E_{eff} \sim (E_1-E_2) l\sim
(E_1-E_2) b^2$ (see (\ref{corrl},\ref{Eeff},\ref{loc})). Thus the only
form of the correlation function consistent with the universality
condition (\ref{corrl}) is
\bq\label{corex}
\overline{Tr \fr{1}{E_1-H}  
          Tr \fr{1}{E_2-H}}_{c} \sim \fr{N}{b(E_1-E_2)^{3/2}}
\left( 1+ \fr{const}{b\sqrt{E_1-E_2}} + ... \right) \ \ \ .
\ee
However, this expression evidently contradicts to the simple estimate
of any of the diagrams of fig.~10.

Thus, we have to choose between the two scenarios. First, as it is
seen from the naive estimates the $\sim 1/b$ corrections to the
correlation function may be much more singular than it is
expected from the universality (\ref{corex}). In this case the
$\sim 1/b$ contributions blow up at $\Delta E$ much larger than
the effective energy level spacing $\Delta E_{eff}$ (\ref{Eeff})
which will be the indication of some new physics at the
intermediate energies $\Delta E_{eff} \ll \Delta E \ll V$~.

In the second scenario the universality (\ref{corrl},\ref{corex}) do
takes place and all the additional singularities cancel each other in
the sum of different diagrams.

In fact the aim of this section is to demonstrate explicitly that at
least the $\sim 1/b$ corrections to the correlation function do not
violate the universality (\ref{corrl},\ref{corex}) and all the
additional singularities vanish after a huge cancellation between the
diagrams of fig.~10.

The analogous cancellation between the high order corrections to the
correlation functions has been previously observed for usual $N\times
N$ random matrices by J.Verbaarschot et
al. \cite{Verb1,Verb2}. 

Consider first the most singular diagrams of fig.~10a. The calculation
of the corresponding contribution to the correlation function of two
logarithms (\ref{LLL}) has much in common with the calculation of the
$\sim 1/b^2$ correction to the Green function at the edge
(\ref{L2H0}-\ref{GH}). The main difference is that now two of the
double-link chains are accompanied by the factor $(V^2
G_0(E_1)G_0(E_2))^p=e^{-\lambda p}$~, while the third chain is
associated with the oscillating contribution $(V^2 G_0^2(E_{1,2}))^k
=\exp(\pm 2ik\phi -k\lambda)$~. It is convenient to brought together
into one expression the both diagrams of fig.~10a
\begin{eqnarray}\label{f10a}
\delta\big( \overline{LL}\big)_a 
&=& \fr{2N}{b^2} \fr{1}{2} \sum_{p_1,p_2>0}
\, \sum_{-\infty <k<\infty} 
\fr{e^{2i\phi k
-\lambda (p_1+p_2+|k|)}}{(2\pi)^d (p_1 p_2 |k|)^{d/2}} \left(
\fr{1}{p_1} +\fr{1}{p_2} +\fr{1}{|k|} \right)^{-d/2} =
\nonumber \\ 
&=& \fr{N}{b^2}\sum \fr{e^{2i\phi k
-\lambda (p_1+p_2+|k|)}}{(2\pi)^d (p_1 p_2
+|k|(p_1 + p_2))^{d/2}} \ \ \ \ . 
\end{eqnarray}
Naively summation over $p_1$ and $p_2$ here gives the singularity
$\sim \lambda^{-2+d}$ (and the additional factor $\lambda^{-2}$ will
appear after differentiation with respect to $E_1$ and $E_2$ which
should be done in order to get the correlation function of two Green
functions). However, at least this leading singularity should
disappear after summation over $k$.  For example, for $d=0$ one has
$\sum_k e^{2i\phi k -\lambda |k|} \sim \lambda$ (even more trivial
example of the same kind is $\sum (-1)^k =0$). For $d\ne 0$ some
cancellation also should take place at least for large $p_{1,2}$
which, on the other hand, are responsible for the ``naive''
singularity of (\ref{f10a}). Thus again one may see that large $k$ or
long chains on the diagram (continuum limit) turn out to be important.

In order to treat this cancellation explicitly it is convenient
to divide the contribution of fig. 10a into two parts
\begin{eqnarray}\label{f10aa}
\delta\big( \overline{LL}\big)_a 
 &=& A+B \ \ \ \ \ , \nonumber \\
A &=& \fr{N}{b^2(2\pi)^d} \sum_{p_1,p_2>0} \, \sum_{-\infty <k<\infty} 
e^{2i\phi k -\lambda (p_1+p_2+|k|)} \left\{ \fr{1}{(p_1 p_2
+|k|(p_1 + p_2))^{d/2}} - \right. \nonumber \\
&-& \left. \fr{1}{p_1^{d/2}(p_2+|k|)^{d/2}}
-\fr{1}{p_2^{d/2}(p_1+|k|)^{d/2}} \right\} \ \ \ \ , \\
B &=& \fr{2N}{b^2(2\pi)^d} \left[ \sum_{p>0} \fr{e^{-\lambda
p}}{p^{d/2}}\right] \sum_{p'>0} \, \sum_{-\infty <k<\infty} 
\fr{e^{2i\phi k -\lambda (p'+|k|)}}{(p'+|k|)^{d/2}}  \ \ \ .
\nonumber 
\end{eqnarray}
Here in the part $A$ the contribution with large both $p_1$ and $p_2$
will be suppressed due to summation over $k$, while the contribution
with say $p_1\ll p_2$ will be suppressed due to simple cancellation of
the two terms in the figure brackets in (\ref{f10aa}). In the second
part $B$ summations over $p$-s are factorized and only summation over
$p'$ suffers from the cancellation due to oscillations.

In the similar way one may write down the contribution of naively
less singular diagrams of fig. 10b
\begin{eqnarray}\label{f10b}
\delta\big( \overline{LL}\big)_b
 &=& \fr{2N}{b^2(2\pi)^d}\left[ \sum_{p>0} \fr{e^{-\lambda
p}}{p^{d/2}}\right] \left\{ \sum_{k \ge 0} e^{(2i\phi
-\lambda)k}  \sum_{q>0} \fr{e^{(2i\phi
-\lambda)q}}{q^{d/2}} + (\phi \rightarrow -\phi) \right\}
= \nonumber \\ 
&=& \fr{2N}{b^2(2\pi)^d}\left[ \sum_{p>0} \fr{e^{-\lambda
p}}{p^{d/2}}\right] \left\{ \sum_{-\infty <k<\infty}
\, \sum_{-|k|<p'\le 0} \fr{e^{2i\phi k
-\lambda |k|}}{(p' + |k|)^{d/2}} \right\}  \ \ \ .
\end{eqnarray}
The situation is further simplified if one combines this
contribution with the most singular part $B$ (\ref{f10aa}) of the
fig. 10a diagrams. After simple change of variables $p' 
\rightarrow q=p'+|k|$ one gets
\begin{eqnarray}\label{f10bB}
\delta\big( \overline{LL}\big)_b
 +B &=& \fr{2N}{b^2(2\pi)^d}\left[ \sum_{p>0} \fr{e^{-\lambda 
p}}{p^{d/2}}\right] \times \\
&\,&
 \sum_{-\infty <k<\infty}
 \, \sum_{q> 0} e^{2i\phi k -\lambda |k|}
 \fr{1}{q^{d/2}} \exp\{ -\lambda(q-|k|)\theta(q-|k|)\} 
 \ \ \ . \nonumber
\end{eqnarray}
Here $\theta(x)=0$ for $x<0$ and $\theta(x)=1$ for $x>0$. Now the
sum over $k$ may be found exactly
\begin{eqnarray}\label{f10bBB}
\delta\big( \overline{LL}\big)_b 
+B = \fr{2N}{b^2(2\pi)^d}\left[ \sum_{p>0} \fr{e^{-\lambda 
p}}{p^{d/2}}\right] \lambda \,  \sum_{q> 0}
\fr{\cos (2q\phi) e^{-\lambda q}}{q^{d/2}2\sin^2\phi} \ \ .
\end{eqnarray}
We see that this contribution as a function of $\lambda$
turns out to be as singular as the leading order result
(\ref{LLL},\ref{Altsh}) but is suppressed like $1/b$ and
therefore should be neglected.

Thus, let us consider the only surviving contribution $A$ from
(\ref{f10aa}). For simplicity consider the band matrices only
($d=1$). After differentiation with respect to $E_1$ and $E_2$ one
gets 
\begin{eqnarray}\label{dGG}
\delta (\overline{ Tr G_1 Tr G_2}) &=& \fr{N}{b^2 2\pi (4V^2-E^2)} \sum
e^{2i\phi k -\lambda (p_1+p_2+|k|)} (p_1+p_2+|k|)^2 \\
&\times&  \left\{ \fr{1}{\sqrt{p_1 p_2
+|k|(p_1 + p_2)}} - 
 \fr{1}{\sqrt{p_1(p_2+|k|)}}
-\fr{1}{\sqrt{p_2(p_1+|k|)}} \right\} \ \ \ . \nonumber
\end{eqnarray}
As we will see both $p_1$ and $p_2$ in this sum effectively turns out
to be large. Therefore, in order not to get exponentially small result
one has to consider the singular in $k$ contributions in the sum. 
These singularities naturally appear due to $|k|$ in (\ref{dGG}).
The following simple identity shows how one may utilize this $\sim |k|$
behavior:
\bq\label{fourier} 
\sum |k| e^{2i\phi k -\lambda |k|}
= \sum |k| e^{2i\phi k -\lambda |k|} f(k) = \fr{-1}{2\sin^2\phi} \ \ \
, 
\ee
where $f(k)$ is any smooth and slow function of $k$ and
$f(0)=1$. Taking into account that $2V\sin \phi
=\sqrt{4V^2-E^2}$ one finds from (\ref{dGG},\ref{fourier})
\bq\label{res}
\delta (\overline{ Tr G_1 Tr G_2}) =\fr{4NV^2}{b^2(4V^2-E^2)^2}
\sum_{p_1,p_2} \fr{p_1+p_2}{2\pi \sqrt{p_1p_2}} e^{-(p_1+p_2)\lambda}
= \fr{-N}{b^2} \fr{2 V^2}{4V^2-E^2} \fr{1}{\Delta E^2} \ \ .
\ee
Finally, the smoothed density-density correlation function for
band matrices takes the form
\bq\label{corrf}
K(E_1-E_2)= \fr{N}{l}
\fr{\overline{\rho(E_1) \rho(E_2)_{c}}}{
\overline{\rho(E_1)} \ \overline{\rho(E_2)} } =
\fr{-1}{2} \left( \fr{\Delta E_{eff}}{ 
|E_1-E_2|}
\right)^{3/2}\left\{ 1+2\left( \fr{\Delta E_{eff}}{
|E_1-E_2|}
\right)^{1/2}+ ... \right\} \ \ \ ,
\ee
where $\Delta E_{eff} = N [2\pi l \overline{\rho(E)}]^{-1}$ and
$l=b^2(1-E^2/4V^2)$ in accordance with
(\ref{corrl},\ref{loc},\ref{corex}). For convenience we have added one
factor $(2\pi)^{-1}$ into the definition of $\Delta E_{eff}$ compared
to (\ref{Eeff}).

To conclude this section let us remind again that we consider only the
smoothed correlation functions. If one would like to compare our
equation (\ref{corrf}) with the result of numerical matrix
diagonalization, the ``experimental'' result should be averaged with
some smooth weight function. For example it may be
\bq\label{exp}
K_{smooth}(\Delta E)= \int K_{exp}(\Delta E+\kappa x) 
e^{-x^2} \fr{dx}{\sqrt{\pi}} \ \ \ ,
\ee
where $\Delta E_{eff} \ll \kappa \ll \Delta E$.

\section{Spatially nonhomogeneous examples}\label{sec:6}

The quantities which we have tried to calculate up to now -- the
density of states and density-density correlation function are
generally considered for the usual $N\times N$ random matrices. In
this section we would like to consider the two quantities which are
specific for band matrices and never appear for the usual $N\times N$
ones.

\begin{figure}
\epsfxsize=8.4 truecm
\epsfysize=7 truecm
\centerline{\epsffile{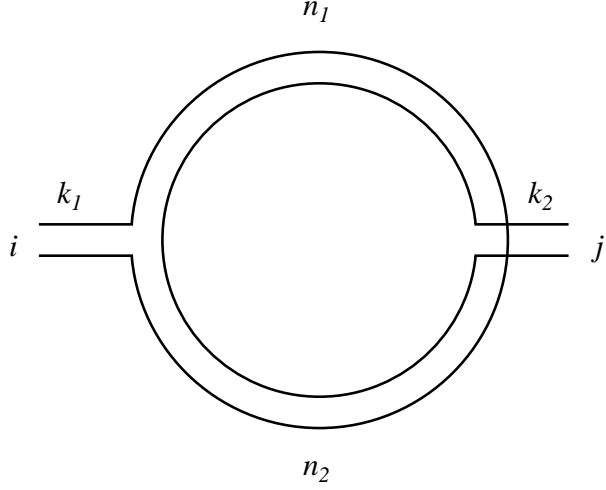}}
\caption{Spatially nonhomogeneous correlation function 
$\overline{ \rho(E_1,i) \rho(E_2,j)}$.}
\end{figure}

The first example will be the correlation function of local density of
states $\rho(E,i)$ for different energies and different vector indices
($1d$-lattice sites) $i$ and $j$. The corresponding dressed Feynman
diagram is shown in the fig. 11.  Again one should calculate first the
log--log correlation function. After differentiation with respect to
$E_1$ and $E_2$ and taking the imaginary part of the Green functions
the correlation function takes the form
\begin{eqnarray}\label{rhoij}
& &\overline{ \rho(E_1,i) \rho(E_2,j)_{c}}=
\fr{1}{\pi^2 b^2} G_0(E_1) G_0(E_2) \\
& & Re \sum_{k_{1,2}\ge 0}
\sum_{n_1+n_2>0} \fr{(G_0(E_1)V)^{2k_1} (G_0(E_2)V)^{2k_2}
(G_0(E_1)G_0(E_2)V^2)^{n_1 +  n_2}}{\sqrt{(k_1+k_2)(n_1+n_2)
+n_1 n_2}} \nonumber \\
& & \ \ \ \ \fr{1}{2\pi} \exp \left\{  -\fr{1}{2b^2}
\fr{(i-j)^2}{k_1+k_2 +\fr{n_1 n_2}{n_1+n_2}} \right\} \ \ \ .
\nonumber 
\end{eqnarray}
Here neither $n_1$ nor $n_2$ could be negative. This equation is
further simplified if one takes into account that effectively
$k_{1,2}\ll n_{1,2}$. Therefore the $k_{1,2}$ may be neglected in the
exponent and in the square root in the denominator and the sum over
$k_1$ and $k_2$ reduces to the simple geometrical progression.
Finally, the summations over $n_1$ and $n_2$ factorize and the
correlation function takes rather simple form
\bq\label{rhoijf}
\overline{ \rho(E_1,i) \rho(E_2,j)_{c}} = 
\fr{1}{\pi^2 b^2
(4V^2-E^2)} Re \ \left[ \sum_n \fr{e^{-\lambda n}}{\sqrt{2\pi n}}
\exp\left\{ -\fr{(i-j)^2}{2nb^2} \right\} \right]^2
\ \ \ .
\ee
Here $\lambda$ is defined by (\ref{LamLL}). One may easily 
investigate for example small and large $i-j$ limits of this 
expression. In terms of universal variables $\Delta E_{eff}$ and $l$ 
(\ref{corrf}) correlation function (\ref{rhoijf}) takes the form
\begin{eqnarray}\label{rhoijun}
\fr{\overline{\rho(E_1,i) \rho(E_2,j)_c}}
{(\overline{\rho(E)}/N)^2}&=&
\fr{1}{2\pi}\left(\fr{i-j}{l}\right)^2 \times \\
& &
Re \ \left[ \int_0^{\infty} \exp\left\{ -\fr{1}{4i} 
\fr{E_1-E_2}{\Delta E_{eff}} \left( \fr{i-j}{l} \right)^2 y^2 
-\fr{1}{2y^2} \right\} dy \right]^2 \ . \nonumber
\end{eqnarray}
Here $Im (E_1-E_2)>0$ and the integral should be squared before 
taking the real part. We have divided $\overline{ \rho(E)}$ by
$N$ in the left hand side of (\ref{rhoijun}) in order to get rid 
of the physically trivial $N$-dependance in the right hand side also.

Finally, the integration over $y$ in (\ref{rhoijun}) may be done
explicitly, which leads to
\begin{eqnarray}\label{rhoijunun}
\fr{\overline{\rho(E_1,i) \rho(E_2,j)_c}}
{(\overline{\rho(E)}/N)^2}&=&
\fr{1}{2} Re \left[ \fr{i\Delta E_{eff}}{E_1-E_2}
\exp \left\{ -\sqrt{2\fr{E_1-E_2}{i\Delta E_{eff}}}
\fr{|i-j|}{l}\right\}\right] = \\
&=& -\fr{1}{2}\left(\fr{i-j}{l}\right)^2
\fr{sin(t) e^{-t}}{t^2} \ \ \ , \nonumber 
\end{eqnarray}
where $t={|i-j|}/{l} \sqrt{{|E_1-E_2|}/{\Delta E_{eff}}}$ . In
particular one may easily examine that after summation over $i$ and
$j$ the equation (\ref{rhoijunun}) leads to the usual correlation
function (\ref{corra}).

Another interesting objects which may be considered easily within our
technic are the finite size band matrices. Naturally the most
interesting case is $b\ll N\le b^2$. Below we describe analytically
the crossover from the band matrix regime (\ref{Altsh}) to
Wigner-Dyson regime in the asymptotics of smoothed density-density
correlation function for finite size band matrices.

Consider the periodic $N\times N$ band matrices. The statistical 
properties of this Gaussian ensemble are again defined by the 
second moment (\ref{band}), but now the function $F$ takes the
form
\bq\label{Fn}
F(i,j)=\sum_{n=-\infty}^{+\infty} f(i-j+nN) \ \ \ .
\ee
Here $f(k)=f(|k|)$ vanishes for $k>b$ just like $F$ in 
(\ref{band}). The parameters $V$ and $b$ (strength of the 
interaction and width of the band) are now defined as
\bq\label{Vn}
V^2= \sum_{j=1}^N F(i,j)=\sum_{-\infty}^{+\infty} f(j) \ \ \ ,
\ \ \ b^2 =\fr{\sum n^2 f(n)}{V^2} \ \ \ ,
\ee
which is the natural generalization of (\ref{defb}).  Also the analog
of the equation (\ref{recur}) for two link chain has the form
\begin{eqnarray}\label{Psinn}
\Psi_{n+1}(i)&=&\sum_{j=1}^N F(i,j)\Psi_{n}(j)
=\sum_{j=-\infty}^{+\infty} f(i-j) \Psi_{n}(j) \ \ , \\
& & \Psi_{n}(j+N) \stackrel{\rm def}{=} \Psi_{n}(j) \ \ \ .
\nonumber 
\end{eqnarray}
The solution of this equation for sufficiently large $n$ 
(and for $b\ll N$) reads
\bq\label{Psinnf}
\Psi_{n}(i)= \fr{V^{2n}}{b\sqrt{2\pi n}} \sum_k \exp\left\{ -
\fr{(i-kN)^2}{2nb^2} \right\} \ \ \ .
\ee
The leading order spherical Green function (\ref{G0}) due to 
(\ref{Vn}) (see also discussion before the equation (\ref{G0}))
is not changed.
Therefore, the trivial modification of (\ref{Altsh}) gives 
\bq\label{rhorhon}
\overline{ \rho(E_1) \rho(E_2)_{c}} = \fr{N}{\pi^2 b
(4V^2-E^2)} Re \ \sum_{p=1}^{\infty} \sum_{k=-\infty}^{\infty}
\sqrt{\fr{p}{2\pi}} e^{-\lambda p} \exp\left\{- \fr{1}{2p} 
\left(
\fr{kN}{b}\right)^2\right\} \ \ ,
\ee
where $\lambda$ again is defined by (\ref{LamLL}). This expression may
be further simplified in two limiting cases.  If $N$ is small, namely
$N \ll b/\sqrt{\lambda}$, one may replace summation over $k$ by
integration. In this case the equation (\ref{rhorhon}) reduces to the
usual Wigner--Dyson correlation function
$\overline{\rho(E_1)\rho(E_2)_c} \sim 1/(E_1-E_2)^2$. If $N \gg
b/\sqrt{\lambda}$ only $k=0$ survive in the second sum and
(\ref{rhorhon}) coincides with the pure band matrix result
(\ref{corra}).

Also it may be convenient to use the ``physical'' variables:
localization length $l=b^2(1-E^2/4V^2)$ and effective interlevel
spacing $\Delta E_{eff}=N [2\pi l\overline{\rho (E)}]^{-1}$ (note that
$\Delta E_{eff}$ does not depend on $N$ because $\overline{\rho
(E)}\sim N$), as it was done in (\ref{corrf}) and (\ref{rhoijun}). Now
one has instead of (\ref{rhorhon})
\begin{eqnarray}\label{rrnun}
\fr{\overline{\rho(E_1)\rho(E_2)_c}}{(\overline{\rho(E)})^2} 
&=& \fr{1}{\sqrt{8\pi}} \fr{l}{N} Re  
\int_0^{\infty} \sum_{k=-\infty}^{\infty} x^2 dx \exp
\left\{ -\fr{1}{4i} \fr{E_1-E_2}{\Delta E_{eff}} x^2
-\left(\fr{kN}{l}\right)^2 \fr{1}{2x^2} \right\} 
= \\
&=&-\fr{l}{2N} 
\left(\fr{\Delta E_{eff}}{|E_1-E_2|}\right)^{3/2}
\left( 1-\tau \fr{d}{d\tau}\right)
\fr{{\rm sh}(\tau)+\sin(\tau)}{{\rm ch}(\tau)-\cos(\tau)}
\ \ \ \ ,\nonumber
\end{eqnarray}
where $\tau= \fr{N}{2l}\sqrt{{|E_1-E_2|}/{\Delta E_{eff}}}$ . In
particular for $\tau \ll 1$ or $\tau \gg 1$ this equation corresponds
to the usual Wigner--Dyson or band matrix (\ref{corrf}) results.

\section{Conclusions}\label{sec:7}

Random matrix models are usually expected to describe some universal
and very general features of complicated quantum systems. Therefore on
the one hand, one and the same very simple model may be associated
with a variety of physical systems. On the other hand, this model will
be generally able to explain only simplified versions of a real
complicated problem, say properties of only very small metallic
grains.  In this paper we have considered banded random matrices which
at least formally seem to be much closer to the real physical
systems. For example, realistic Hamiltonians in the shell model for
complicated atom \cite{Flamb} and atomic nuclei \cite{Zelev} were
shown to have a banded structure. Also, being a good example of
quasi-$1d$ quantum system random band matrices are expected to depict
adequately properties of electrons in thick wires
\cite{Mirlin,Pichard,efetov} (see also \cite{IWZ} there the mapping of
the Hamiltonian for disordered wire onto random banded block-diagonal
matrix was done explicitly).

Technically our work was stimulated in part by the successful
application of matrix models for calculation of partition function of
$2d$-quantum gravity. Just like in $2d$-gravity we have found a
critical behaviour at the edge of spectrum for band matrices (Section
{\bf 3}) and for lattice Hamiltonian with random hopping (Section {\bf
4}). Unfortunately it may be not so easy to find a physical system
whose global spectral properties will be described by the random band
matrices with their almost semicircle density of states. Nevertheless,
the interest in investigation of the edge behavior and the tails of
spectral density have been demonstrated in recent papers
\cite{Chirikov95,Felix}.

In almost all applications of random matrix theory one is interested
in local characteristics of spectrum such as correlations of very
close or even neighbouring energy levels. In Section {\bf 5} we have
calculated the asymptotics of two-point correlation function of the
density of states (\ref{Altsh},\ref{corra}) which is in agreement with
the result of ref. \cite{Altshuler} for energy level correlations in
disordered metals. Moreover, together with the hypothesis of
universality of spectral correlations \cite{Aronov} our result
(\ref{corra}) allowed to estimate the localization length (\ref{loc})
for random band matrices and this calculation seems to be much less
complicated than known from literature \cite{Mirlin}.

On the other hand, we are able to calculate only the asymptotics (plus
corrections) of the correlation function which in principle should not
necessarily manifest universality. The universal behaviour of
(\ref{corra}) shows that there are only two different energy scales in
the model the global width of the energy zone $\Delta E=4V$ and the
effective inter-level splitting $\Delta E_{eff}$
(\ref{Eeff},\ref{corrf}). The hypothesis of universality finds further
support in the calculation of the first correction to the two-point
correlation function (\ref{corrf}). To the best of my knowledge it is
the first calculation of subleading corrections for quasi-$1d$
systems. However, the calculation of the correction also shows a
serious drawback of our perturbative approach. The accurate result
(\ref{res},\ref{corrf}) was found only after the huge two-step
cancellation. One may speculate that this is the price for working
very far from the region of convergence of the initial series in
$H/E$. Nevertheless, these cancellations show that it will be
extremely difficult to reach the region $E_1-E_2 \sim \Delta E_{eff}$
in our approach.

Finally, in the last section we have found the asymptotics of the
local density of states two-point correlation function (\ref{rhoijun})
as well as the usual two-point correlator for finite size quasi-$1d$
system (\ref{rrnun}). These relatively simple analytical calculations
demonstrate again the usefulness of diagrammatic approach for
investigation of such nontrivial systems as random band matrices.

{\bf Acknowledgements.} \\ Author is thankful to J.~Ambj{\o}rn,
B.~V.~Chirikov, Y.~V.~Fyodorov, F.~M.~Izrai\-lev, Yu.~A.~Makeenko,
A.~D.~Mirlin, D.~V.~Savin, V.~V.~Sokolov and O.~P.~Sushkov for useful
discussions. The figures have been drown by L.~F.~Hailo.


\end{document}